\documentclass[twocolumn]{aastex61}
\usepackage{amsmath}
\usepackage{CJK}

\newcommand{\SE}{{\rm SE}}
\newcommand{\CJ}{{\rm CJ}}
\newcommand{\mps}{{\rm m~s^{-1}}}
\newcommand{\mjup}{M_{\rm J}}

\shorttitle{The Super Earth-Cold Jupiter Relations}
\shortauthors{Zhu \& Wu}

\begin{document}
\begin{CJK*}{UTF8}{gbsn}

\title{The Super Earth-Cold Jupiter Relations}

\author{Wei~Zhu (祝伟)}
\affil{Canadian Institute for Theoretical Astrophysics, University of Toronto, 60 St. George Street, Toronto, ON M5S 3H8, Canada}
\correspondingauthor{Wei Zhu}
\email{weizhu@cita.utoronto.ca}

\author{Yanqin Wu (武延庆)}
\affil{Department of Astronomy and Astrophysics, University of Toronto, 50 St. George Street, Toronto, ON M5S 3H4, Canada}

\begin{abstract}
We report relations between inner ($<1$ au) super Earths (planets with mass/radius between Earth and Neptune) and outer ($>1$ au) giant planets (mass $>0.3~\mjup$, or cold Jupiters) around Sun-like stars, based on data from both ground-based radial velocity (RV) observations and the \emph{Kepler} mission. We find that cold Jupiters appear three times more often around hosts of super Earths than they do around field stars. Given the prevalence of the super Earth systems, their cold Jupiters can account for nearly all cold Jupiters. In other words, cold Jupiters are almost certainly ($\sim90\%$) companied by super Earths.
A few corollaries follow: (1) around metal-rich ([Fe/H]$>0.1$) stars, the fraction of super Earths with cold Jupiters can rise to 60\% or higher; (2) the inner architecture can be strongly impacted by the outer giant and we report some observational evidence for this; (3) planetary systems like our own, with cold Jupiters but no super Earths, should be rare ($\sim1\%$). 
The strong correlation between super Earths and cold Jupiters establish that super Earths and cold Jupiters do not compete for solid material, rather, they share similar origins, with the cold Jupiter formation requiring a somewhat more stringent condition. Lastly, we propose a few immediate observational tests of our results, using ground-based RV observations and ongoing/planned space missions.
\end{abstract}                                     


\section{Introduction} \label{sec:introduction}

Planets with masses/radii between Earth and Neptune, commonly called super Earths, have unknown origins. Such planets do not present in our Solar system, and are not predicted by models of planet formation \citep{IdaLin:2004,Mordasini:2009}. However, the \emph{Kepler} mission has demonstrated that they are ubiquitous in the Galaxy: they appear around nearly $30\%$ of all Sun-like stars \citep{Zhu:2018} and can be formed around stars with a broad range of stellar metallicities \citep{Udry:2006,Buchhave:2012} and masses \citep{Dressing:2013}. Some recent ideas have been proposed to explain their origin, but a consensus has yet to be reached (see recent reviews by \citealt{Morbidelli:2016} and \citealt{Schlichting:2018}).

Observationally, there are several pathways toward a better understanding of the super Earth population, each complimentary to the others. First, mass and/or radius measurements of these planets provide insights into their bulk compositions and evolution paths \citep[e.g.,][]{Wu:2013,Marcy:2014,Hadden:2017,Owen:2017}. Detections of or constraints on super Earth atmospheres can provide information on their compositions \citep[e.g.,][]{Kreidberg:2014,Tsiaras:2016}. In addition, one can study the correlations (if any) between super Earth occurrence rates and host star properties to infer the requirements on their birth environment \citep[e.g.,][]{Wang:2015,Mulders:2015,Zhu:2016,Petigura:2018}.  Last but not least, one can investigate the relations between the super Earth population and other planet populations, such as the giant planets.

Depending on their separations from the hosts, giant planets ($\gtrsim0.3~M_{\rm J}$) can be further divided into different categories, with the cold giants ($P\gtrsim1~$au, hereafter cold Jupiters) overwhelmingly dominating the overall population. These latter planets appear around $10\%$ of Sun-like stars \citep{Cumming:2008}, and exhibit a strong dependence on stellar metallicities \citep{Santos:2001,Fischer:2005}. They are thought to be formed in the outer regions, with the planetary core gradually built up through planetesimal accretion, followed by run-away gas accretion after the core has reached a critical mass \citep{Pollack:1996}. While some details remain unsolved \citep[e.g.,][]{Helled:2014,Morbidelli:2016}, our understanding of the cold Jupiter population, both observationally and theoretically, is far better than our understanding of super Earths. In this work, we hope to leverage our knowledge of cold Jupiters toward solving the mystery of super Earths.

Proposed ideas for the formation of super Earths can be divided into three categories: \textit{in situ} formation, formation-then-migration, and migration-then-assembly. They predict (sometimes implicitly) either correlation or anti-correlation between super Earths and cold Jupiters. If the super Earths are formed \textit{in situ} out of local material, then the required protoplanetary disk must be very massive \citep{Chiang:2013}. With the nominal surface density profile, the outer disk should be more massive than the minimum-mass solar nebulae (MMSN, \citealt{Weidenschilling:1977,Hayashi:1981}) and are therefore more likely to form cold Jupiters. On the formation-then-migration side, two models explicitly predict an anti-correlation between the two populations: \citet{IdaLin:2010} suggested that super Earths and cold Jupiters should preferentially form around metal-poor and metal-rich stars, respectively; \citet{Izidoro:2015} proposed that the early-formed cold Jupiters would act as barriers to the inward migration of super Earths. 
Two representatives of the migration-then-assembly scenario are by \citet{Hansen:2012} and \citet{Chatterjee:2014}. In the former, super Earths are conglomerated in the inner region without gas assist (similar to the theory of terrestrial formation), although the material is assumed to come from the outer region. In the latter (also called ``inside-out formation''), planestesimals migrate inward and collect in local pressure bumps in the gas disk. This triggers formation of a super Earth, which subsequently moves the pressure bump outward, prepping for the formation of another planet. Both scenario do not make explicit predictions on cold Jupiters, but since the solid material is sourced from the Jovian region, one naively expects an anti-correlation.

In this work, we use planetary systems from ground-based RV and space-based \emph{Kepler} observations to reveal the super Earth-cold Jupiter relationship. Specifically, we derive the frequencies of cold Jupiters (/super Earths) in systems already with super Earths (/cold Jupiters) in Section~\ref{sec:correlation}. We carry out an observational test of the proposed relations in Section~\ref{sec:multiplicity}. Discussions of our results are given in Section~\ref{sec:discussion}, and in Section~\ref{sec:predictions} we list several ways to test and/or refine the super Earth-cold Jupiter relations in the near future.

\section{Cold Jupiters Coexist with Super Earths} \label{sec:correlation}

\begin{figure*}
\epsscale{1.}
\plotone{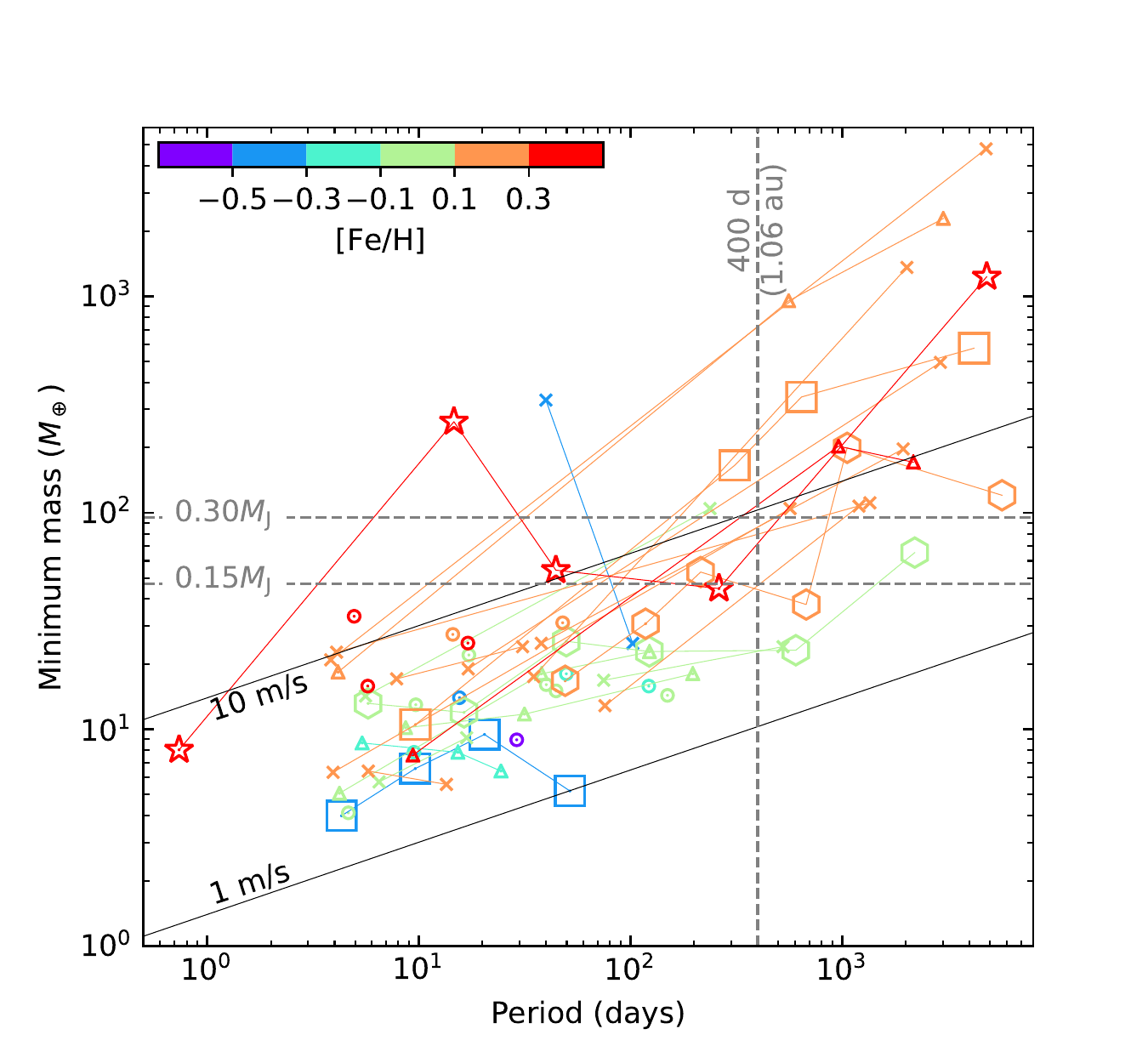}
\caption{Minimum masses ($m\sin{i}$) and orbital periods of the RV planets in our sample. Colors represent the stellar metallicities ([Fe/H]), and symbols represent the number of planet detections in the same system: circle for 1-planet, cross for 2-planet, triangle for 3-planet, square for 4-planet, asterisk for 5-planet, and hexagon for 6-planet, respectively. Planets with the same host are connected with solid lines. The horizontal gray dashed lines indicate the two mass thresholds for our initial sample selection, and the vertical gray dashed line indicates the boundary between inner and outer planetary systems. The black solid lines denote two characteristic RV semi-amplitudes. Most cold Jupiters induce RV semi-amplitude larger than that by super Earths. Therefore, the RV series capable of revealing super Earths are mostly capable of revealing cold Jupiters, as long as the RV time-spans are long enough.
\label{fig:rv_sample}}
\end{figure*}

We first constrain the frequency of systems with cold Jupiters given that there is already at least one super Earth in the system. Mathematically, this is represented by the conditional probability $P(\CJ|\SE)$, with ``CJ'' and ``SE'' standing for cold Jupiter and super Earth for short, respectively. We will then combine this conditional probability with the absolute occurrence rates $P(\CJ)$ and $P(\SE)$, to infer $P(\SE|\CJ)$. The two conditional probabilities, $P(\CJ|\SE)$ and $P(\SE|\CJ)$, fully describe the correlation between the two planet populations.

\subsection{Conditional Probability $P(\CJ|\SE)$} \label{sec:cond_prob_cj}

\begin{figure*}
\epsscale{1.1}
\plotone{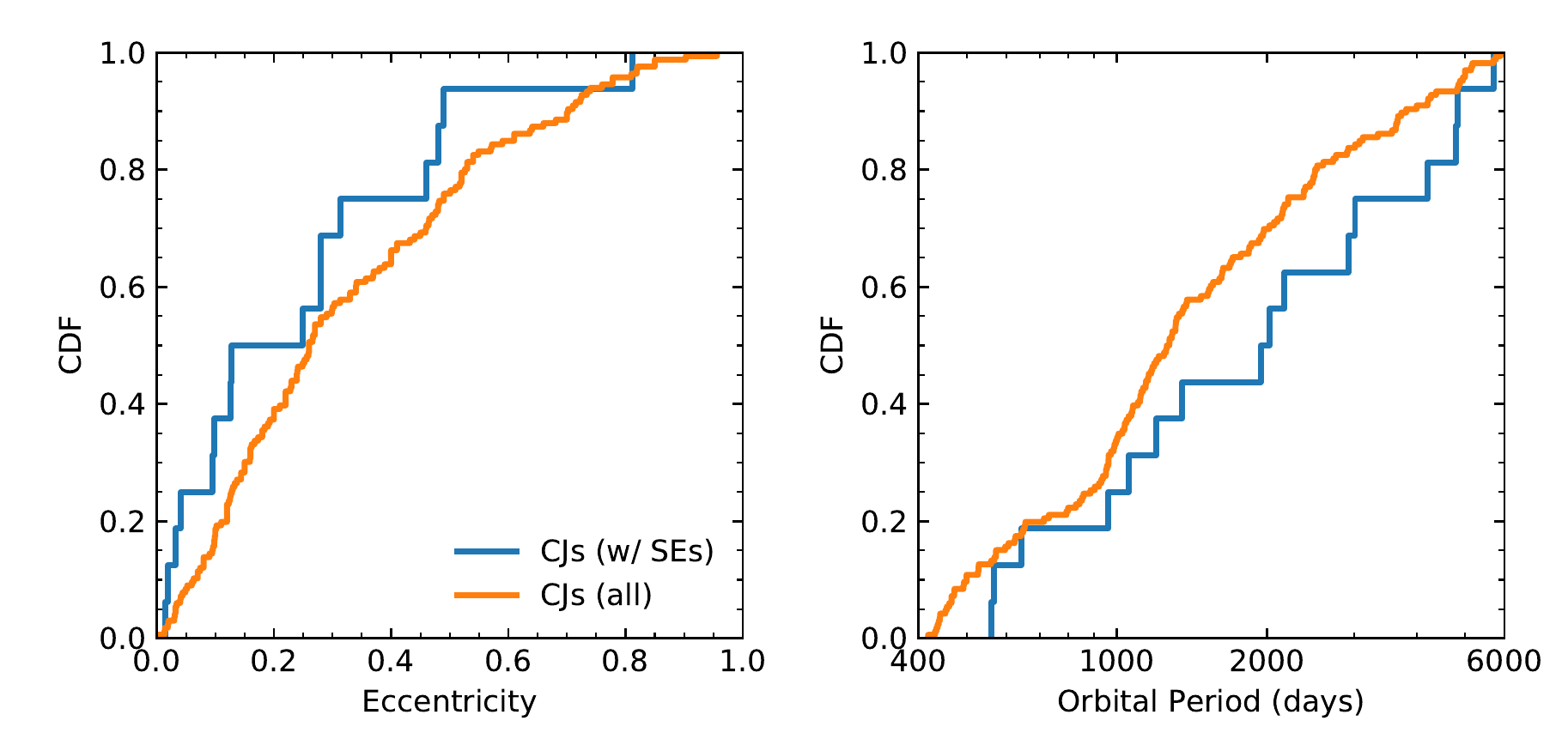}
\caption{Cumulative distributions of orbital eccentricities (left panel) and periods (right panel) of cold Jupiters (CJs). Blue curves are for CJs with super Earth (SE) companions (i.e., CJs in our sample), and orange curves are for all CJs found by RV observations. Data are taken from NASA Exoplanet Archive. Two-sample KS tests give $p=0.29$ and $0.31$ for the eccentricity distributions and period distributions between two CJ samples, respectively, suggesting that the CJs with SEs are statistically the same as the overall CJ population.
\label{fig:comparisons}}
\end{figure*}

\begin{figure}
\epsscale{1.1}
\plotone{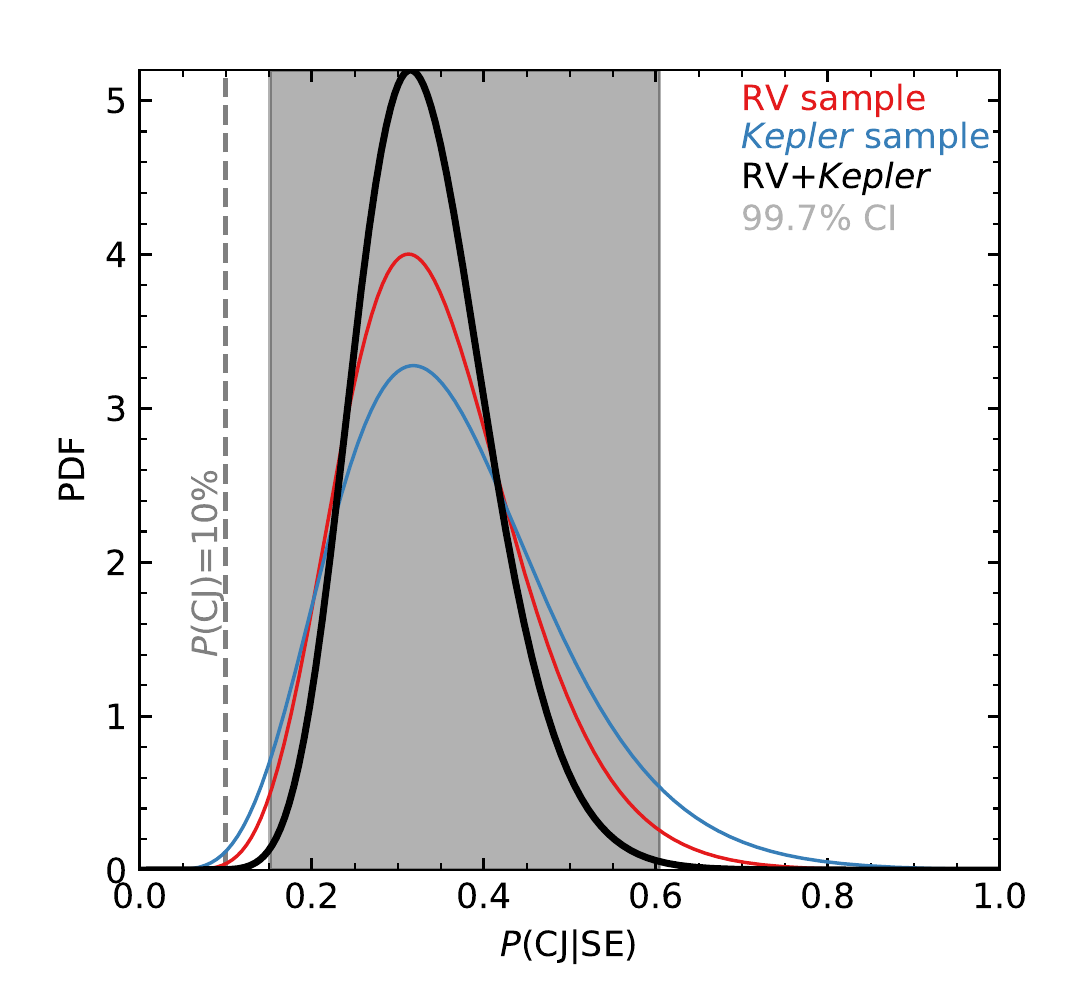}
\caption{Posterior distributions of the conditional probability $P(\CJ|\SE)$, constructed from the ground-based RV sample, the \emph{Kepler} sample, and the joint sample. For the ground-based RV sample, a super Earth is defined by $m\sin{i}<20~M_\oplus$ and $P<400~$d. The shaded region represents the $99.7\%$ confidence interval (CI) centered on the median of the posterior distribution from the joint sample. The gray dashed line indicates the value if there is no correlation between super Earth and cold Jupiter.
\label{fig:posterior}}
\end{figure}

We constrain $P(\CJ|\SE)$, first using systems discovered by ground RV surveys. This is a reasonable approach, because the RV data capable of revealing super Earths are almost certainly capable of revealing cold Jupiters as long as the RV time-span is long enough. As shown in Figure~\ref{fig:rv_sample}, a cold Jupiter, defined as a planet with mass $m\sin{i}>0.3~\mjup$ (i.e., $95~M_\oplus$) and orbital period $P\gtrsim$1 yr, induces an RV amplitude larger than (or at least comparable to) that induced by a super Earth, even though the latter is closer to the star.
\footnote{This is no longer true for low-mass, very far-away ($P>20$ yrs) cold Jupiters. However, statistical studies suggested that such planets should be very rare \citep{Bryan:2016}.}
Therefore, the efficiency of detecting the cold Jupiter is nearly 100\% in a system with a super Earth already detected, provided that the star has been monitored long enough (i.e., at least half the orbital period of the cold Jupiter).

We start from the confirmed planetary systems in the NASA Exoplanet Archive.
\footnote{Data used in this study were obtained from \url{https://exoplanetarchive.ipac.caltech.edu} on 2018 March 4.}
We first remove all claimed planet detections with RV semi-amplitude $K<1~\mps$, given their high refutation probability (e.g., $\alpha$ Cen Bb, \citealt{Rajpaul:2016}).
Then for a system to be included in our sample, we require that
\begin{enumerate}
\item All the planets are first detected by RV;
\item There is at least one planet with $m\sin{i}<47~M_\oplus$ and $P<400$ d;
\item The host is Sun-like, defined as $T_{\rm eff}$ in the range $4700-6500$ K and $\log{g}>4.0$.
\footnote{In cases where these parameters are not given in NASA Exoplanet Archive, we extract them from the original discovery papers.}
\end{enumerate}
The first criterion excludes systems with planets first detected by transit and then followed up by RV, because we want to separate the pure-RV sample and the \emph{Kepler} sample. The second criterion is our most generous definition for super Earths (see Table~\ref{tab:fractions}), which are spaced in mass by at least a factor of two below the cold Jupiters. These criteria lead to a sample of 39 systems with a total of 82 planets. We show these planets in Figure~\ref{fig:rv_sample}, and provide detailed information for all of them in Appendix~\ref{appendix}. Almost all systems have RV time-spans longer than four years. So a null detection almost certainly means no cold Jupiters out to 8-yr orbit. There are two systems with RV time-spans less than four year, both being single systems. They are included nevertheless, and excluding them would further strengthen our conclusion.

Among our sample of 39 super Earth systems, 12 contain cold Jupiters. To obtain statistical results on cold Jupiters, we should first make sure that these cold Jupiters share similar properties as the overall cold Jupiter population. Figure~\ref{fig:comparisons} shows the cumulative distributions of eccentricities and orbital periods of cold Jupiters in our sample as well as all the cold Jupiters included in NASA Exoplanet Archive. The two-sample Kolmogorov-Smirnov (KS) test gives $p=0.29$ and $0.31$, respectively, suggesting that they are drawn from the same underlying population. Of the 12 systems with cold Jupiters, five have at least two Jovian-mass ($>0.3~\mjup$) planets. The multiplicity rate ($42\pm19\%$) is also broadly consistent with the multiplicity rate of overall cold Jupiters ($\sim28\%$, \citealt{Wright:2009}).

Given the nearly unity detection efficiency of the cold Jupiters, the conditional probability $P(\CJ|\SE)$ is simply given by the fraction of systems with cold Jupiters in our RV sample. We compute this under different definitions of super Earths, as well as including or excluding systems with warm Jupiters ($m\sin{i}>95~M_\oplus$ and $10~$d$<P<100~$d), and report the results in Table~\ref{tab:fractions}. We find that $P(\CJ|\SE)$ is $\sim$30\% in nearly all cases, except where a super Earth is given by the most stringent definition ($m\sin{i}<10~M_\oplus$), and thus the sample size is reduced significantly to 10.

\begin{deluxetable}{lcc}
\tablecaption{Conditional probabilities $P(\CJ|\SE)$ under different definitions of super Earths. The second case is our nominal definition, and the results from it are marked in bold.
\label{tab:fractions}}
\tablehead{
\colhead{Super Earth (SE)} & \colhead{$P(\CJ|\SE)$} & \colhead{$P(\CJ|\SE)$} \\
\colhead{definitions} & \colhead{(incl. WJ)$^a$} & \colhead{(excl. WJ)$^a$}}
\startdata
$m\sin{i}<47~M_\oplus,~P<400~$d & 12/39 & 11/37 \\
$m\sin{i}<20~M_\oplus,~P<400~$d & \textbf{10/32} & \textbf{9/31} \\
$m\sin{i}<20~M_\oplus,~P<100~$d & 10/30 & 9/29 \\
$m\sin{i}<10~M_\oplus,~P<100~$d & 3/11 & 2/10 \\
\enddata
\tablecomments{$^a$ A warm Jupiter (WJ) is a planet with $m\sin{i}>95~M_\oplus$ (i.e., $0.3~M_{\rm J}$) and $10~$d$<P<100~$d, and a cold Jupiter (CJ) is a planet with $m\sin{i}>95~M_\oplus$ and outside 1 au.}
\end{deluxetable}

\citet{Cumming:2008} constrained the demographics of giant planets within 2000 days. Applying their giant planet distribution function to our period range ($400-8000$ days), one obtains that 10\% of Sun-like stars should have at least one cold Jupiter, or $P(\CJ)=10\%$.
\footnote{\citet{Cumming:2008} only included the most detectable planet around any star in deriving the fraction of stars with planets. This technique tends to overestimate the fraction, as explained in \citet{Zhu:2018}. Taking into account the overestimation in $P(\CJ)$ would further strengthen our results.}
Our finding that $P(\CJ|\SE)\approx30\%$ therefore indicates that cold Jupiters appear three times more often around super Earth hosts than they do around field Sun-like stars.

Although Table~\ref{tab:fractions} suggests that the conditional probability $P(\CJ|\SE)$ is largely independent of the mass definition for super Earths, the super Earths detected by RV are systematically more massive than typical super Earths. To see whether our result also holds for lower-mass super Earths, we construct a sample of super Earth systems found by the transit method from the \emph{Kepler} mission, and with RV follow-up observations for at least one year. Here we also only include Sun-like ($4700$ K$<T_{\rm eff}<6500$ K and $\log{g}>4.0$) stars. These super Earths are so low in mass that in many cases only upper limits can be derived from RV. We therefore define a super Earth as a planet with radius $R_{\rm p}$ in the range $1-4~R_\oplus$ and orbital period $P<400$ d. From NASA Exoplanet Archive, we find 22 such systems. We also provide detailed information for these systems in Appendix~\ref{appendix}. Among them, seven have been reported to have a cold Jupiter companion. Therefore, $P(\CJ|\SE)=32\%$, if the detection efficiency is also close to unity. Lower detection efficiencies would further increase this conditional probability and therefore strengthen our later results. So in summary, this independent \emph{Kepler} sample, comprising mostly of lower-mass super Earths, confirms the result from the ground-based RV sample that cold Jupiters occur three times more often around hosts of super Earths.

We further quantify the significance of the above claim in the following way. For any given value of $P(\CJ|\SE)$ ranging from zero to unity, we compute the likelihood that $m$ out of the $n$ super Earth systems contain cold Jupiters, where we have $(m,n)=(10,32)$ and $(7,22)$ for the ground-based RV sample and the \emph{Kepler} sample, respectively. According to Bayes theorem, this likelihood serves as the posterior probability of $P(\CJ|\SE)$ at the given value. The full posterior distributions of $P(\CJ|\SE)$ constructed in this way are shown in Figure~\ref{fig:posterior} for the ground-based RV sample, the \emph{Kepler} sample, and the combined sample. In particular, the posterior distribution based on the combined sample indicates that $P(\CJ|\SE)>P(\CJ)$ at $>99.7\%$ confidence level. In other words, cold Jupiters appearing more frequently in super Earth systems is detected at $>3\sigma$ level.

\subsection{Conditional Probability $P(\SE|\CJ)$} \label{sec:cond_prob_se}

We now derive the other conditional probability $P(\SE|\CJ)$, a number that is perhaps more physically revealing. In principle one can use a similar method as we just did for $P(\CJ|\SE)$, but given the more difficult technical requirement in detecting super Earths by RV, this approach is currently not possible. In particular, current RV instrument is not sensitive to the bulk of the super Earth population (Figure~\ref{fig:rv_sample}).

We therefore choose an alternative approach utilizing the Bayes theorem
\begin{equation} \label{eqn:bayes}
P(\SE) \times P(\CJ|\SE) = P(\CJ) \times P(\SE|\CJ)\ .
\end{equation}
The conditional probability we seek can be obtained if we also know $P(\CJ)$ and $P(\SE)$. The former is established by \citet{Cumming:2008} to be $\sim$10\% (see previous section), and here we focus on the latter quantity.

As discussed in \citet{Zhu:2018}, the fraction of stars with planets is harder to constrain than the average number of planets per star, because of the intrinsic multiplicity. As a result, $P(\SE)$, the fraction of Sun-like stars with at least one super Earth, has not been correctly and rigorously constrained. Here, we take $P(\SE)=30\%$, where the latter is the occurrence rate of \emph{Kepler}-like planetary systems as determined by \citet{Zhu:2018}. This occurrence rate refers to planets detectable by the \emph{Kepler} mission, namely, planets that are mostly larger than Earth in size and that orbit shortward of 400 days. We believe this is a reasonable approximation for $P(\SE)$ for the following reasons.

First, the majority of planets discovered by \emph{Kepler} are super Earths, with a minor smattering of hot Jupiters and warm Jupiters. Excluding these latter planets would reduce $P(\SE)$ only slightly from 30\%. Second, our $P(\SE)$ refers to that in the RV sample, while the 30\% of \citet{Zhu:2018} refers to the \emph{Kepler} sample. Fortunately, the metallicity distributions of the \emph{Kepler} stars and the RV stars are very similar \citep{Guo:2017}, and the very minor difference introduces negligible effect, because the super Earth occurrence rate has a weak dependence on stellar metallicity \citep{Udry:2006,Buchhave:2012,Wang:2015,Zhu:2016}.

With $P(\CJ|\SE)=32\%$, $P(\SE)=30\%$, and $P(\CJ)=10\%$, Equation~(\ref{eqn:bayes}) yields $P(\SE|\CJ)=96\%$. Therefore, systems with cold Jupiters almost certainly have super Earths. We defer a detailed discussion on the implications in Section~\ref{sec:discussion}. Instead, we address a few possible caveats here.

First, although the uncertainty on the derived $P(\SE|\CJ)$ remains significant, the strong correlation between super Earths and cold Jupiters is secure mostly because of the well determined $P(\CJ|\SE)$ (see Figure~\ref{fig:posterior}). Ignoring the asymmetric probability distribution of $P(\CJ|\SE)$ as well as the uncertainties on $P(\CJ)$ and $P(\SE)$, we have $P(\SE|\CJ)=90\pm20\%$. The systematic overestimation in $P(\CJ)$, as explained in Section~\ref{sec:cond_prob_cj}, can further enhance the already strong correlation. Regardless of the actual value of $P(\SE|\CJ)$, our result indicates that the non-correlation case, $P(\SE|\CJ)=P(\SE)$, and particularly the anti-correlation case, $P(\SE|\CJ)=0$, can be securely excluded.

If only high-metallicity ([Fe/H]$>0.1$) systems are included, we find that out of the 29 systems of our combined sample, 17 have cold Jupiters, so $P(\CJ|\SE)=17/29\approx60\%$, twice higher than the overall sample. Meanwhile, the value of $P(\CJ)$ also rises to $\sim$20\% \citep{Santos:2001,Fischer:2005}, while the value of $P(\SE)$ changes only marginally because of its weak dependence on stellar metallicity. As a result, we obtain that $P(\SE|\CJ)\approx90\%$. This is broadly consistent with the previous result (96\%), and excludes the possibility that the high value of $P(\SE|\CJ)$ is due to the metallicity effect. Furthermore, the above result also shows that cold Jupiters and super Earths almost occur concomitantly in the metal-rich systems.

The value of $P(\SE|\CJ)$ cannot be exactly 100\%. The size of our combined sample is large enough to reliably constrain $P(\CJ|\SE)$ and therefore $P(\SE|\CJ)$, but not large enough to include the rare cases, such as hot Jupiter (HJ) systems and our Solar system. First, with very rare exceptions (e.g., WASP-47b, \citealt{Becker:2015}), hot Jupiters usually do not have close-by small companions \citep{Steffen:2012}, and yet $\sim$50\% of hot Jupiters have cold Jupiter companions \citep{Knutson:2014}. With $P({\rm HJ})\approx 1\%$ \citep{Wright:2012}, this gives $P({\rm HJ}|\CJ)\approx5\%$ and therefore sets an upper limit of $P(\SE|\CJ)\lesssim95\%$. This value can be further reduced by systems like our own, which has cold Jupiters but no super Earths. However, the amount of reduction is not yet known.

Interestingly, we can inversely infer the prevalence of Solar system analogs, given our result on $P(\SE|\CJ)$. The probability of having cold Jupiter but no super Earth (no-SE) in a given system is given by
\begin{equation}
P({\rm no{\text-}SE},\CJ) = [1-P(\SE|\CJ)]P(\CJ)\approx1\%\ .
\end{equation}
In the above evaluation we have adopted $P(\SE|\CJ)\approx90\%$ and $P(\CJ)=10\%$. So, while only a small fraction (10\%) of Sun-like stars have cold Jupiters, an even smaller fraction (1\%) of Sun-like stars have cold Jupiters but no super Earths. We note that because of the uncertainties on the input parameters $P(\SE|\CJ)$ and $P(\CJ)$, the resulting fraction is only accurate to order-of-magnitude level. Nevertheless, our result implies that planetary systems like our own are rare.

\section{Cold Jupiter Reduces Inner Multiplicity?} \label{sec:multiplicity}

\begin{figure*}
\epsscale{1.1}
\plotone{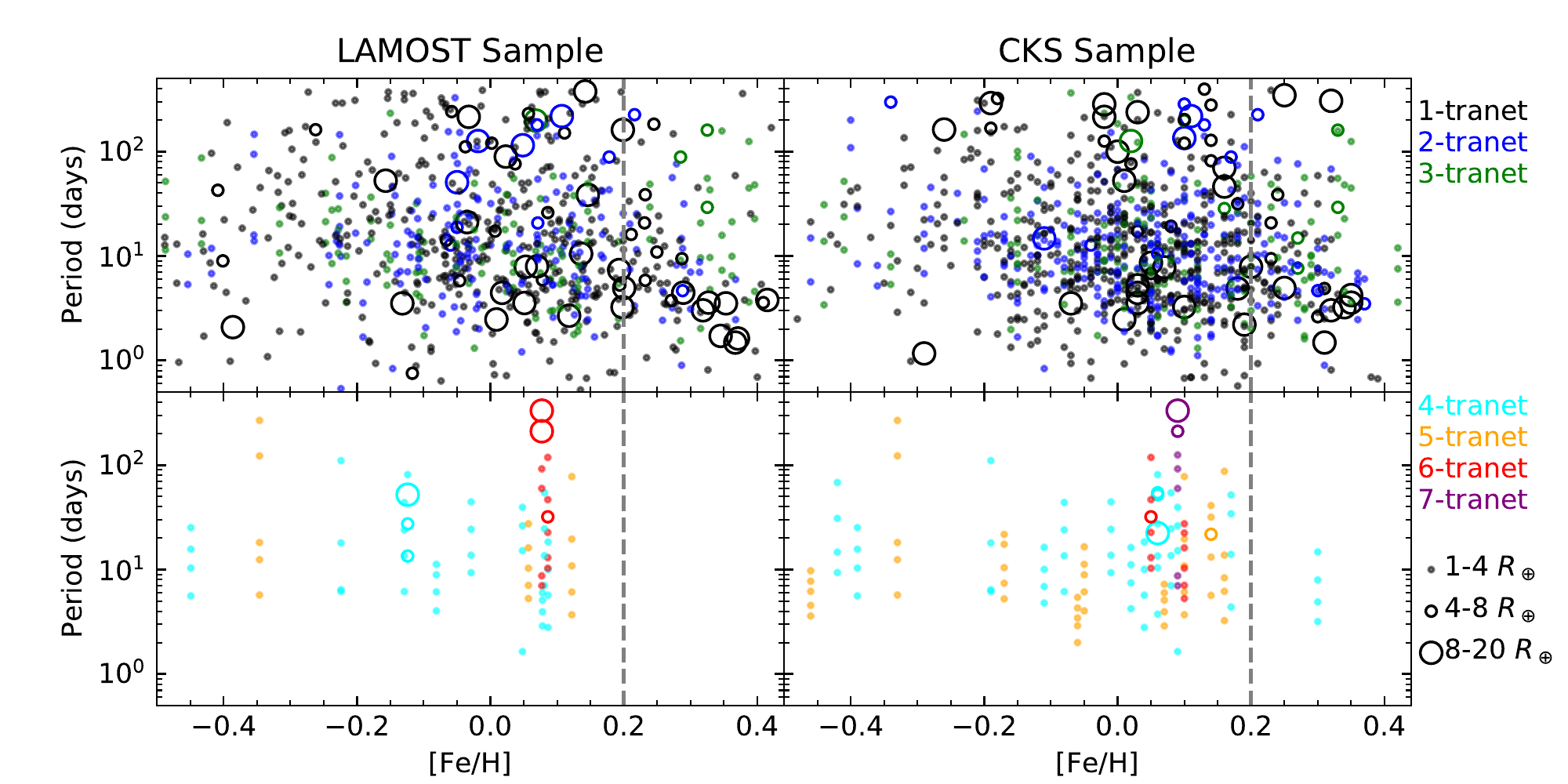}
\caption{The \emph{Kepler} planets in LAMOST (left panels) and CKS (right panels) samples. The top panels illustrate the planets in low-multiplicity systems (i.e., no more than three transiting planets), and the bottom panels the planets in high-multiplicity systems (i.e., at least four transiting planets). The location of individual planet is determined by the stellar metallicity ([Fe/H]) and the planetary orbital period. The size of the symbol denotes the radius of the planet. However, for clarity, planets are divided into three classes: 1-4 $R_\oplus$, 4-8 $R_\oplus$, and 8-20 $R_\oplus$. For both samples, marginal evidence is seen that there is a deficit of high-multiple systems in the high-metallicity regime. The [Fe/H]$=0.2$ lines are marked out for guidance.
\label{fig:multiplicity}}
\end{figure*}

Cold Jupiters typically have significant eccentricities ($e\sim0.3$, Figure~\ref{fig:comparisons}). Therefore, if coexisting with inner systems, they are expected to disturb the inner planetary systems dynamically \citep{Matsumura:2013,Huang:2017,Mustill:2017,Hansen:2017,Becker:2017,Pu:2018}. One therefore expects that cold Jupiters should be largely incompatible with heavily packed multi-planet systems, in particular if these inner systems are already near the edge of dynamical instability \citep{Pu:2015}. As a result, the systems with cold Jupiters should have on average fewer planets within 400-day orbit than the ones without cold Jupiters. This is consistent with our \emph{Kepler} sample (Table~\ref{tab:kepler}): systems with and without cold Jupiter detections have on average 1.4 and 2.5 planets within 400-day orbit, respectively. However, this sample is small and may suffer from detection bias, so we look for additional evidence that supports the above prediction.

Although it is not yet possible to detect cold Jupiters for the majority of \emph{Kepler} super Earth systems, we can study this issue by taking proxies. While the frequency of cold Jupiters in all super Earth systems is $P(\CJ|\SE)\approx30\%$, this frequency rises to $\sim$60\% in metal-rich systems ([Fe/H]$>$0.1, see previous section). If dynamical disturbances by cold Jupiters are significant, we expect to see a more severe impact in metal-rich systems. This motivates us to look for an anti-correlation between multiplicity of the inner systems and stellar metallicity. Here we also adopt the observed transit multiplicity as a proxy for the underlying multiplicity of the inner system.

We employ two samples of well charaterized \emph{Kepler} planets, one from the LAMOST
\footnote{The Large Sky Area Multi-Object Fiber Spectroscopic Telescope \citep{Cui:2012}.}
survey of the \emph{Kepler} field \citep{Dong:2014,Decat:2015,Ren:2016} as refined in \citet{Zhu:2018}, and the other from the magnitude-limited CKS
\footnote{The California-Kepler Survey \citep{Petigura:2017,Johnson:2017}. We only include stars with \emph{Kepler} magnitude $K_p<14.2$ as suggested by \citet{Fulton:2017}.}
survey. There are significant overlaps between the two samples, but we choose not to merge them, considering their similar but not the same metallicity scales. We illustrate both samples in Figure~\ref{fig:multiplicity}. Specifically, we separate the planetary systems into low-multiplicity ($k\le3$) and high-multiplicity ($k\ge4$) groups based on the observed transit multiplicity $k$, and planets in each group are assigned locations based on the stellar metallicities and planetary periods. As shown in Figure~\ref{fig:multiplicity}, the outermost planets in the high-multiple systems typically have orbital periods $P\gtrsim100$ days and thus semi-major axes $a\gtrsim0.4$ au, making them vulnerable to outer giant companions. For example, with the minimum mass ($M_{\rm p}=100~M_\oplus$) and separation ($a_{\rm p}=1$ au) for the outer cold giant, any planet with mass $m<(a/a_{\rm p})^3 M_{\rm p}/0.3=21~M_\oplus$ would be disturbed \citep{Lai:2017}.

While low-multiplicity systems show up at all metallicity values, there appears to be a shortage of high-multiple systems around metal-rich stars. With 16 and 25 high-multiplicity systems at low metallicity ([Fe/H]$<0.2$), one expects to see 2.7 and 4.2 high-multiplicity systems at high metallicity ([Fe/H]$>0.2$) in the LAMOST and CKS samples, respectively, if the planet multiplicity is independent of the stellar metallicity. Instead, we see zero and one, respectively. The only high-multiple system in the metal-rich environment has all four planets within 20 days, and thus it can survive perturbations from most cold Jupiters \citep{Lai:2017}. Even with this exceptional case included, there is a deficit of high-multiple systems in the high-metallicity regime at the 93\% confidence level.
\footnote{This is the probability to have at least one such system in the LAMOST sample (or two such systems in the CKS sample).}
Combining the two samples does not increase the confidence level because of the significant overlap between the two samples.

While being only a marginal detection, the above result is consistent with our expectation. A larger sample is required to firmly establish it.

\section{Discussion} \label{sec:discussion}

\begin{figure}
\epsscale{1.2}
\plotone{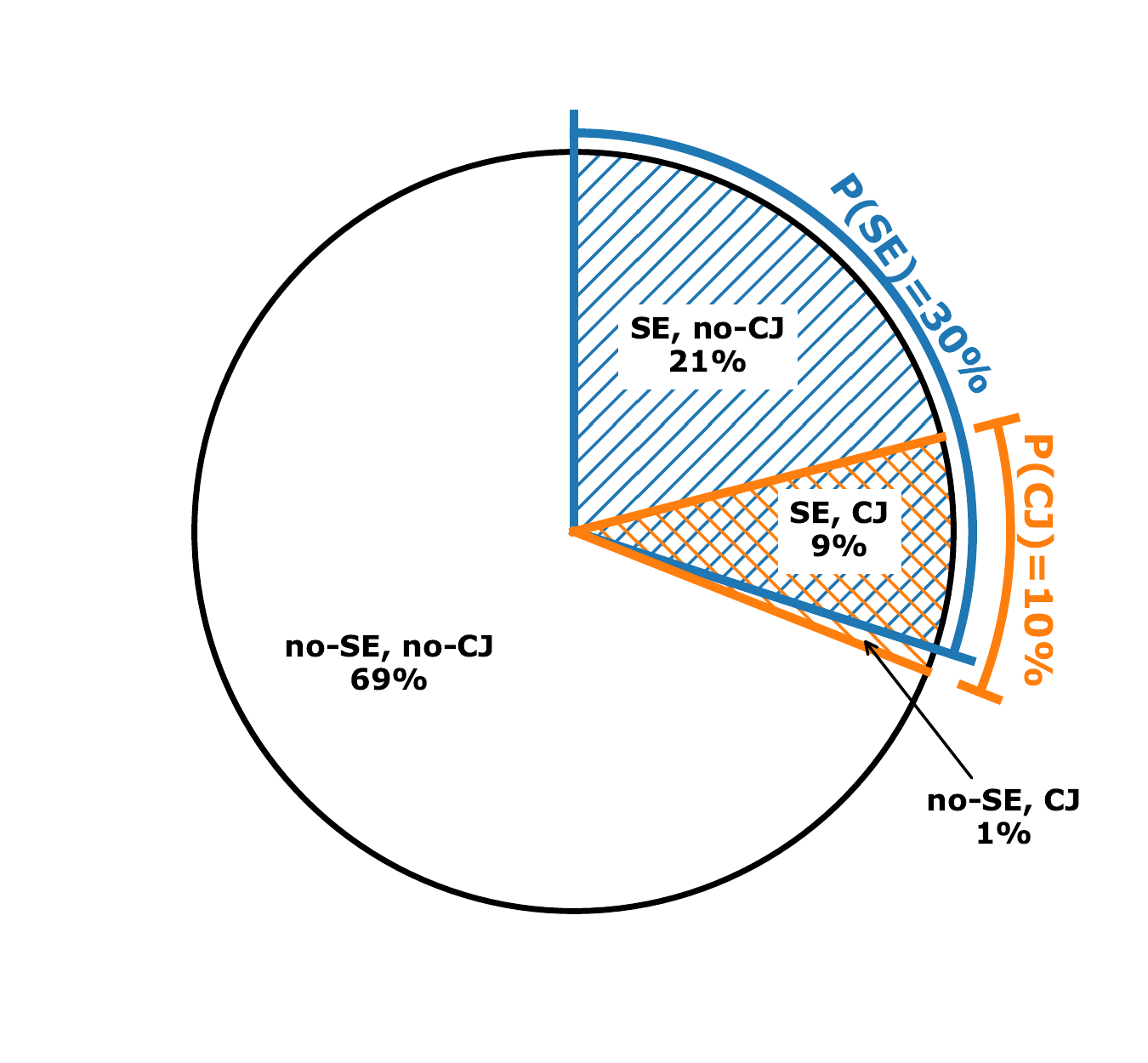}
\caption{An illustrative Venn-Pie diagram that summarizes the percentages for different types of planetary systems. Here ``SE'' stands for super Earth, ``CJ'' for cold Jupiter. The $1\%$ probability for planetary systems like the Solar system (with no SE, but CJ) is only accurate to order-of-magnitude level. Minority systems like those containing hot Jupiters ($\sim1\%$) are not here.} 
\label{fig:illustration}
\end{figure}

We report the following super Earth-cold Jupiter relations for Sun-like stars:
\begin{enumerate}
\item $P(\CJ|\SE)=32\%\pm8\%$: stars with super Earths have $\sim3$ times higher cold Jupiter fraction, compared to field stars ($\sim10\%$). Because of the giant planet-metallicity correlation, this fraction increases with the stellar metallicity and rises to 60\% for stars with [Fe/H]$>0.1$.
\item $P(\SE|\CJ)=90\%\pm20\%$
\footnote{This uncertainty is derived by assuming that the uncertainty on $P(\CJ|\SE)$ dominates and that it is symmetric. The upper bound at unity is not imposed.}:
stars with cold Jupiters almost certainly have super Earths. Because of the weak dependence of super Earth occurrence rate on stellar metallicity, this fraction is largely independent of the stellar metallicity.
\end{enumerate}
The first relation is derived from two complementary and independent samples: stars in the Solar neighborhood that have been known to have (high-mass) super Earths by RV surveys, and the stars in the \emph{Kepler} field that have been known to have (low-mass) super Earths by transit. In both samples, the precision and time-span of RV observations are sufficient to detect the cold Jupiters, and therefore the apparent ratio of cold Jupiter hosts directly yields $P(\CJ|\SE)$. The consistency between the RV sample and the \emph{Kepler} sample also indicates that this conditional fraction is independent of the super Earth mass.

The second relation is derived from the Bayes theorem (Equation~(\ref{eqn:bayes})), with the known absolute fractions of super Earth systems ($\sim30\%$, \citealt{Zhu:2018}) and cold Jupiter systems ($\sim10\%$, \citealt{Cumming:2008}). In this way, we avoid the non-trivial calculation of detection efficiency of super Earths in cold Jupiter systems. We do not take into account hot Jupiter systems, most of which do not have super Earths \citep{Steffen:2012}. The inclusion of such rare cases imposes an upper limit $P(\SE|\CJ)\lesssim95\%$.

{A recent study by \citet{Barbato:2018} searched in 20 cold Jupiter systems for inner ($P<150$ days) planets with $m\sin{i}$ in the range $10-30~M_\oplus$, and found zero detection. Given the small sample size, their zero detection result is statistically consistent with our findings. With the relatively massive super Earths denoted as ``mSE'' one would expect that $P({\rm mSE}|\CJ)=P(\CJ|{\rm mSE}) P({\rm mSE})/P(\CJ) \approx 30\%$. Here we have used the results that $P({\rm mSE})\approx P(\CJ)=10\%$
\footnote{\citet{Barbato:2018} used 38.8\% for $P({\rm mSE})$. However, this fraction is the fraction of stars having planets with masses up to $30~M_\oplus$, rather than in the range $10-30~M_\oplus$. See Table 1 of \citet{Mayor:2011}.}
and that $P(\CJ|{\rm mSE})\approx P(\CJ|\SE)$ (see Figure~\ref{fig:rv_sample}). With an average sensitivity of $50\%$, one therefore would expect to have 3 low-mass planet detections in a sample of 20 cold Jupiter systems. The zero detection result is then consistent at 95\% confidence level. If only the lower-mass range ($10-20~M_\oplus$) is considered (so as to be consistent with our definition of super Earth), the expected number of detections is even lower. This explains why \citet{Barbato:2018} did not detect any such planets.
}

Based on our results, we illustrate in Figure~\ref{fig:illustration} a break-down of planetary varieties into four kinds, based on the absence/presence of super Earths and cold Jupiters, respectively. Majority of the systems do not contain super Earths or cold Jupiters, although they may still contain other types of planets that are less readily detectable by current transit or RV techniques, such as our own terrestrial planets.

There are two very direct consequences of our super Earth-cold Jupiter relations. First, because typical cold Jupiters have significant orbital eccentricities, they ought to have imprints on the architecture of the inner planetary systems. By investigating the planetary systems with spectroscopic measurements from LAMOST and CKS, we do find evidence that heavily packed systems appear less often around metal-rich stars than they do around metal-poor ones. Although larger samples are required to strengthen the significance of this effect, it appears consistent with and in favor of our super Earth-cold Jupiter relations.

The second consequence of our relations is about the prevalence of planetary systems like our own, which has cold Jupiters but no super Earths. Given the low probability of having cold Jupiters and a similar probability of not having super Earths if cold Jupiters are already present, Solar system analogs are very rare ($\sim1\%$). However, such a low occurrence rate may be related to our stringent definition of a Solar system analog. In particular, our Earth straddles the boundary between super Earths and sub-Earths.

We now turn to the implications of our results on planet formation theories.  The cores of cold Jupiters should form well before the gas is depleted from the protoplanetary disks \citep{Pollack:1996}. The existence of hydrogen envelopes on many of the super Earths \citep{Wu:2013,Marcy:2014}, in particular those that have not suffered the fate of photo-evaporation \citep{Owen:2013,Lopez:2013,Fulton:2017}, also dictates their early formation. As such, it is not surprising to find that the two types of planets are correlated. Our results are most naturally explained in a formation scenario where many planetary cores form early in the gas disks, both inside and outside of the ice line, and only the most massive ones undergo run-away gas accretion to become giant planets. Unfortunately, no such a scenario is readily available at the moment, so we turn our eyes to those proposed models.

Two theories explicitly predicted an anti-correlation between super Earths and cold Jupiters.  \citet{IdaLin:2010} proposed that super Earths and cold Jupiters (or more generally, giant planets) form in different metallicity environments: when metallicity is high, there is enough solid material to form the giants before gas depletion; otherwise, super Earths are formed. On the other hand, \citet{Izidoro:2015,Izidoro:2017} argued that super Earths are formed further out in the gas disks and then migrated inward, and that cold Jupiters, if present, should act as barriers to this inward migration. Both of these theories are excluded by the positive correlation that we discover.

Instead, the strong correlation between super Earths and cold Jupiters suggests that these two planet populations do not directly compete for solid material. This, taken at its face value, disfavors theories that invoke large-scale and substantial migration of solid material across protoplanetary disks, in the form of dust particles, pebbles, or proto-planets.

\section{Predictions} \label{sec:predictions}

We propose below multiple ways to further test and/or refine our super Earth-cold Jupiter relations, all achievable within the next few years.

First, for stars hosting known cold Jupiters, RV observations can be conducted to detect the presence of shorter period super Earths. With typically a few $M_\oplus$, the bulk of super Earths are barely detectable with even the currently state-of-the-art RV instruments (see Figure~\ref{fig:rv_sample}). Therefore, in order to accumulate a sizeable sample of super Earths to perform meaningful statistics, a fairly large sample of cold Jupiter hosts should be intensively observed.
Alternatively, one can hope to use the next-generation RV instrument that aims to reach $\sim10$ cm s$^{-1}$ precision \citep{Fischer:2016}.

Second, nearly all the cold Jupiter hosts will be monitored by the just-launched Transiting Exoplanet Survey Satellite (TESS, \citealt{Ricker:2015}), and possibly later on by the CHaracterizing ExOPlanet Satellite (CHEOPS, \citealt{Broeg:2013}).  We expect that, for the $\sim 100$ such systems currently known, about a dozen should show transiting super Earths, a factor of three higher than similar systems without cold Jupiters. 

Third, among the over $1000$ transiting super Earths promised to be detected by TESS \citep{Sullivan:2015}, a third of them should be accompanied with cold Jupiters. These planets are relatively easily detectable by ground-based RV follow-ups, or by the GAIA mission through the astrometry method \citep{Perryman:2014}.

\acknowledgements
We would like to thank Songhu Wang, Cristobal Petrovich, Dong Lai, Jiwei Xie, and Subo Dong for helpful discussions. We also thank the anonymous referee for comments on the paper.
This paper includes data collected by the \emph{Kepler} mission. Funding for the \emph{Kepler} mission is provided by the NASA Science Mission directorate.
This research has made use of the NASA Exoplanet Archive, which is operated by the California Institute of Technology, under contract with the National Aeronautics and Space Administration under the Exoplanet Exploration Program.
This paper also uses data from the LAMOST survey.
Guoshoujing Telescope (the Large Sky Area Multi-Object Fiber Spectroscopic Telescope, LAMOST) is a National Major Scientific Project built by the Chinese Academy of Sciences. Funding for the project has been provided by the National Development and Reform Commission. LAMOST is operated and managed by the National Astronomical Observatories, Chinese Academy of Sciences. YW acknowledges funding from NSERC.

\appendix
\section{Super Earth Systems from RV and \emph{Kepler}} \label{appendix}

We provide in Table~\ref{tab:rv_sample} the information of super Earth systems included in our RV sample. These systems are selected from the NASA Exoplanet Archive, following the procedures detailed in Section~\ref{sec:cond_prob_cj}.

We also provide in Table~\ref{tab:kepler} the list of \emph{Kepler} super Earth systems. These systems all contain at least one planet with radius $R_\oplus<R_{\rm p}<4~R_\oplus$ and period $P<400$ days, and they have been followed up by RV for at least one year.

\startlongtable
\begin{deluxetable}{lrcllll}
\tablecaption{RV systems in our sample, listed in the order of increasing [Fe/H]. Planets in individual systems are listed in the order of increasing periods, and cold Jupiters are marked in bold.
\label{tab:rv_sample}}
\tablehead{
\colhead{Host name} & \colhead{[Fe/H]} & \colhead{RV time} & \colhead{Planet} & \colhead{Periods} & \colhead{Reference$^a$} & \colhead{Comments} \\
 & & \colhead{span (years)} & \colhead{$m\sin{i}$ ($M_\oplus$)} & \colhead{(days)} & & }
\startdata
HD 175607 & $-0.62$ &  9.2 & 9.0 & 29.0 & \citet{Mortier:2016} & (1) \\
HD 40307 & $-0.31$ & 10.4 & 4.0, 6.6, 9.5, & 4.3, 9.6, 20.4, & \citet{Diaz:2016} & \\
 &  & & 5.2 & 51.8 &  & \\
HD 4308 & $-0.31$ & 1.9 & 14.0 & 15.6 & \citet{Udry:2006} & \\
$\rho$ CrB & $-0.31$ & 8.0 & 332.1, 25.0 & 39.8, 102.5 & \citet{Fulton:2016} & (2) \\
HD 97658 & $-0.30$ & 5.5 & 7.9 & 9.5 & \citet{VanGrootel:2014} & (3) \\
HD 102365 & $-0.26$ & 12 & 16.0 & 122.1 & \citet{Tinney:2011} & (4) \\
HD 90156 & $-0.24$ & 4.4 & 18.0 & 49.8 & \citet{Mordasini:2011} & \\
HD 7924 & $-0.15$ & 10.2 & 8.7, 7.9, 6.4 & 5.4, 15.3, 24.5 & \citet{Fulton:2015} & (5) \\
HD 42618 & $-0.09$ & 9.1 & 14.4 & 149.6 & \citet{Fulton:2016} & \\
BD-08 2823 & $-0.07$ & 5.0 & 14.4, 104 & 5.6, 237.6 & \citet{Hebrard:2010} & \\
HD 69830 & $-0.05$ & 5.0 & 10.2, 11.8, 18.1 & 8.7, 31.6, 197 & \citet{Lovis:2006} & (6) \\
HD 192310 & $-0.04$ & 6.4 & 16.9, 24.0 & 74.7, 525.8 & \citet{Pepe:2011} & \\
HD 164595 & $-0.04$ & 2.2 & 16.1 & 40.0 & \citet{Courcol:2015} & \\
61 Vir & $-0.01$ & 4.6 & 5.1, 18.2, 22.9 & 4.2, 38.0, 123.0 & \citet{Vogt:2010} & \\
HD 125595 & $0.02$ & 5.7 & 13.1 & 9.7 & \citet{Segransan:2011} & \\
HD 176986 & $0.03$ & 13.2 & 5.7, 9.2 & 6.5, 16.8 & \citet{SM:2017} & \\
HD 156668 & $0.05$ & 4.5 & 4.2 & 4.6 & \citet{Howard:2011} & (7) \\
HD 16417 & $0.06$ & 10.6 & 22.1 & 17.2 & \citet{OToole:2009} & \\
HD 10180 & $0.08$ & 6.7 & 13.2, 12.0, 25.6, & 5.8, 16.4, 49.7, & \citet{Kane:2014} & (8) \\
 &  & & 22.9, 23.3, 65.7 & 122.7, 604.7, 2205 & & \\
HD 177565 & $0.08$ & 4.6 & 15.1 & 44.5 & \citet{Feng:2017} & \\
HD 109271 & $0.10$ & 7.4 & 17.2, 24.2 & 7.9, 30.9 & \citet{LoCurto:2013} & (9) \\
HD 11964 & $0.12$ & 12 & 25.0, \textbf{198.0} & 37.9, \textbf{1945} & \citet{Wright:2009} & \\
HD 34445 & $0.14$ & 18 & 16.8, 30.7, 53.5, & 49.2, 117.9, 214.7, & \citet{Vogt:2017} & \\
 &  &  & 37.9, \textbf{200.0, 120.6} & 676.8, \textbf{1056.7, 5700} & & \\
HD 164922 & $0.16$ & 19 & 12.9, \textbf{107.6} & 75.8, \textbf{1201.1} & \citet{Fulton:2016} & \\
HD 204313 & $0.18$ & 8.5 & 17.6, \textbf{1360} & 34.9, \textbf{2024} & \citet{Diaz:2016} & (10) \\
HD 1461 & $0.19$ & 10.2 & 6.4, 5.6 & 5.8, 13.5 & \citet{Diaz:2016} & (11) \\
HD 219828 & $0.19$ & 14.1 & 21.0, \textbf{4799} & 3.8, \textbf{4791} & \citet{Santos:2016} & (12) \\
HD 103197 & $0.21$ & 6.1 & 31.2 & 47.8 & \citet{Mordasini:2011} & \\
HD 125612 & $0.23$ & 5.5 & 18, \textbf{968, 2286} & 4.2, \textbf{559, 3008} & \citet{LoCurto:2010} & \\
HD 47186 & $0.23$ & 4.3 & 22.8, \textbf{111.4} & 4.1, \textbf{1353.6} & \citet{Bouchy:2009} & \\
HD 215497 & $0.23$ & 5.1 & 6.6, \textbf{104.3} & 3.9, \textbf{567.9} & \citet{LoCurto:2010} & \\
HD 190360 & $0.25$ & 11.8 & 19.1, \textbf{495.8} & 17.1, \textbf{2915} & \citet{Courcol:2015} & \\
HD 179079 & $0.25$ & 4.3 & 27.5 & 14.5 & \citet{Valenti:2009} & \\
HD 160691 & $0.26$ & 7.1 & 10.6, 165.9, \textbf{343.2,} & 9.6, 310.6, \textbf{643.2,} & \citet{Pepe:2007} & \\
 & & & \textbf{576.5} & \textbf{4206} & & \\
HD 99492 & $0.30$ & 18.2 & 25 & 17.1 & \citet{Kane:2016} & (13) \\
55 Cnc & $0.31$ & 23.2 & 8.1, 264, 54.5, & 0.7, 14.7, 44.4, & \citet{Baluev:2015} & (14) \\
 &  & & 44.8, \textbf{1232.5} & 262, \textbf{4825} &  & \\
HD 49674 & $0.31$ & 5.1 & 33.4 & 4.9 & \citet{Butler:2006} & \\
HD 181433 & $0.33$ & 4.8 & 7.5, \textbf{203, 171} & 9.4, \textbf{962, 2172} & \citet{Bouchy:2009} & \\
HD 77338 & $0.35$ & 7.2 & 15.9 & 5.7 & \citet{Jenkins:2013} & \\
\enddata
\tablecomments{$^a$ In cases where multiple references are available, we only provide the most recent ones. \\
(1) A planet candidate with period $1400$ days was reported, but its mass ($34~M_\oplus$) does not qualify for a cold Jupiter. \\
(2) Earlier astrometric studies \citep[e.g.,][]{Reffert:2011} indicated that this system should be face-on and thus the companion b would be a low-mass star. However, this is not confirmed by \citet{Fulton:2016}. \\
(3) The planet was found to transit its host \citep{Dragomir:2013}. \\
(4) The host has an M-dwarf companion at 211 au \citep{Raghavan:2010}. \\
(5) A 6.6-year periodic signal was detected. However, it is likely due to the stellar magnetic activity cycle. \\
(6) A 8500-day long-period trend is seen, but most likely it is induced by the magnetic cycle of the star \citep{Anglada-Escude:2012}. \\
(7) A 2.2-year periodic signal was also reported, but the origin remained unknown. Even if this signal was due to a second planet, the derived mass ($45~M_\oplus$) would not qualify it for a cold Jupiter. \\
(8) Three more candidates were also reported, all with super Earth masses \citep{Lovis:2011,Tuomi:2012}. \\
(9) A 430-day periodic signal was also detected, but it would not qualify for a cold Jupiter even if this signal was due to another planet. \\
(10) A second giant ($1.7~M_{\rm J}$) planet with $P\sim2800$ days reported by \citet{Robertson:2012} was not confirmed. \\
(11) A long-period ($\sim10$ years) periodic signal is detected, but most likely it is due to the magnetic cycle of the star. Even if this could originate from a planet, the planet would not qualify for a cold Jupiter. \\
(12) The cold Jupiter in this system has a mass marginally exceeding $13~M_{\rm J}$. \\
(13) The period ($\sim5000$ days) of the proposed giant ($0.36~M_{\rm J}$, \citealt{Meschiari:2011}) correlates with the stellar activity cycle, and therefore is not confirmed as a planet. \\
(14) The innermost planet (e) transits the host star \citep{Winn:2011}.
}
\end{deluxetable}

\begin{deluxetable*}{lrcccl}
\tablecaption{Kepler super Earth systems with RV observations longer than one year. Systems are listed in the order of increasing stellar metallicities ([Fe/H]).
\label{tab:kepler}}
\tablehead{
\colhead{Name} & \colhead{[Fe/H]} & \colhead{Multiplicity$^a$} & \colhead{Minimum mass$^b$ ($\mjup$)} & \colhead{Period$^b$ (days)} & \colhead{Reference}}
\startdata
Kepler-37 & $-0.32$ & 4 & \nodata & \nodata & \citet{Marcy:2014} \\
Kepler-97 & $-0.20$ & 1 & $>1.08$ & $>789$ & \citet{Marcy:2014} \\
Kepler-93 & $-0.18$ & 1 & $>3$ & $>1460$ & \citet{Marcy:2014} \\
Kepler-10 & $-0.15$ & 2 & \nodata & \nodata & \citet{Dumusque:2014} \\
Kepler-106 & $-0.12$ & 4 & \nodata & \nodata & \citet{Marcy:2014} \\
Kepler-25 & $-0.04$ & 3 & \nodata & \nodata & \citet{Marcy:2014} \\
Kepler-21 & $-0.03$ & 1 & \nodata & \nodata & \citet{LopezMorales:2016} \\
kepler-100 & 0.02 & 3 & \nodata & \nodata & \citet{Marcy:2014} \\
Kepler-89 & 0.02 & 4 & \nodata & \nodata & \citet{Weiss:2013} \\ 
Kepler-96 & $0.04$ & 1 & \nodata & \nodata & \citet{Marcy:2014} \\
Kepler-113 & $0.05$ & 2 & \nodata & \nodata & \citet{Marcy:2014} \\
Kepler-20 & $0.07$ & 6 & \nodata & \nodata & \citet{Buchhave:2016} \\
Kepler-68 & $0.12$ & 2 & $0.95\pm0.04$ & $580\pm15$ & \citet{Marcy:2014} \\
Kepler-131 & $0.12$ & 2 & \nodata & \nodata & \citet{Marcy:2014} \\
Kepler-48 & $0.17$ & 3 & $2.07\pm0.08$ & $982\pm8$ & \citet{Marcy:2014} \\
Kepler-98 & $0.18$ & 1 & \nodata & \nodata & \citet{Marcy:2014} \\
Kepler-99 & $0.18$ & 1 & \nodata & \nodata & \citet{Marcy:2014} \\
Kepler-406 & $0.18$ & 2 & \nodata & \nodata & \citet{Marcy:2014} \\
Kepler-454 & $0.27$ & 1 & $4.46\pm0.12$ & $523.9\pm0.7$ & \citet{Gettel:2016} \\
Kepler-95 & $0.30$ & 1 & \nodata & \nodata & \citet{Marcy:2014} \\
Kepler-407 & $0.33$ & 1 & $12.6\pm6.3$ & $3000\pm500$ & \citet{Marcy:2014} \\
Kepler-94 & $0.34$ & 1 & $9.8\pm0.6$ & $820\pm3$ & \citet{Marcy:2014} \\
\enddata
\tablecomments{$^a$ This is the observed multiplicity within $P<400$ days. Only confirmed planets are included. \\
$^b$ These are the parameters of the detected cold Jupiters.}
\end{deluxetable*}


\end{CJK*}
\end{document}